# A complete Raman mapping of phase transitions in Si under indentation

C. R. Das [a], H. C. Hsu [b], S. Dhara [a, c,*], A. K. Bhaduri [a], B. Raj [a], L. C. Chen [b], K. H. Chen [b,d], S. K. Albert [a], A. Ray [e], and Y. Tzeng [c]

[a] Metallurgy and Materials Group, Indira Gandhi Center for Atomic Research, Kalpakkam-603102, India

[b] Center for Condensed Matter Sciences, National Taiwan University, Taipei-106, Taiwan

[c] Department of Electrical Engineering, Institute for Innovations and Advanced Studies, National Cheng Kung University, Tainan-701, Taiwan.

[d] Institute of Atomic and Molecular Sciences, Academia Sinica, Taipei-106, Taiwan

[e] Department of Physics, Indian Institute of Technology, Kharagpur- 721 302, India

*Abstract*

Crystalline Si substrates are studied for pressure induced phase transformation under indentation at room temperature using Berkovich tip. Raman scattering study is used for the identification of the transformed phases. Raman line as well as area mapping are used for locating the phases in the indented region. Calculation of pressure contours in the indented region is used for understanding the phase distribution. We report here a comprehensive study of all the phases of Si, reported so far, leading to possible understanding of material properties useful for possible electromechanical applications. As a major finding, distribution of amorphous phase in the indented region deviates from the conventional wisdom of being in the central region alone. We present phase mapping results for both Si(100) and Si(111) substrates.

* Corresponding author Email:s_dhara2001@yahoo.com

On leave from Materials Science Division, Indira Gandhi for Centre for Atomic Research, Kalpakkam 603102, India

# 1. Introduction

Intense localized stresses and strains are experienced by materials under a rigid diamond indenter. These high stresses are reported for pressure-induced phase transformations to denser crystalline (c-) and amorphous forms in monovalently [1-3] and covalently bonded compound [4,5] semiconductors along with plastic deformation, leading to recrystallization by dislocation activity in ionic bonded compound [6] semiconductors. Study of materials under pressure is always of immense interest both for electronic and mechanical properties. In addition, the study of transformation to the high pressure phase has technological importance for having control in the precision micro-machining process steps using enhancement of ductility [7,8] in these systems. In these contexts, a complete understanding of phase transformation of Si under high pressure is of renewed interest in the age of ion cutting and ion doping to suit Si based devices, namely, very very large scale integrated (VVLSI) technology and micro-electromechanical (MEMS) or even nano- electromechanical systems (NEMS) systems [9].

In the pressure range up to 40 GPa, diamond-like Si-I (0-11 GPa) phase is reported to transform to a metastable phase of body centered tetragonal β-Sn structured Si-II (11-15 GPa), which on further relaxation transforms to various phases of Si, namely, Si-III (10-0 GPa) having crystallographic structure of body centered cubic with eight atoms at the basis (*bcc*8), and Si-XII (12-2 GPa) rhombohedral (*r*8) structure [2,3,10,11]. Relatively sparse phases of Si-V (primitive hexagonal) ~14 → 40 GPa; and Si-VII (hexagonal close-packed) ~40 GPa are also reported [10]. Amorphization of Si is also reported at the center of various indentors [1-3]. Apart from crystallographic evidences, there are Raman evidences for all these metastable phases of Si leaving Si-II (metallic), Si-V and Si-VII phases [2,3]. However, there is hardly any report of Raman mapping for all these phases. It necessitates a study in this regard when uses of Si based MEMS/NEMS are getting extremely popular [9].

## 2. Experimental

Boron doped *p*-type (100) and (111) oriented Si substrates of 300 μm thickness are used in the present study. The samples are indented using the micro-indenter with a Berkovich diamond indenter (three-sided pyramid with a nominal tip radius of 50 nm and having centerline-to-face angle, $\Psi=65.3^o$). The indentation conditions are as follows: load 50-700 mN; loading-unloading rate 5-10 $mN.s^{-1}$, and holding time 5 s. The optical image of the indentation is taken using microscope with 100X/0.9 objective (Olympus, BXFM). The length of the side of the triangle indentation is about 10 μm. A LabRam HR800 (Jobin-Yvon, France) spectrometer with an automatized XY-table of acquisition is used to record the Raman spectra with excitation wavelength of 632.8nm nm of He-Ne laser. Raman area mapping is also recorded for the spectral region of interest. All the Raman spectra are recorded in backscattering geometry.

## 3. Results and discussion

*3.1 Mechanical Properties*

Typical hardness (Fig. 1) of c-Si is studied at various indentation loads with two different loading rates. Hardness values equivalent to bulk Si (~8.5 GPa) [1] and the value reported using Berkovich indenter (~11.5-12.5 GPa)[12] are also achieved at high loads > 300 mN. As a matter of fact hardness increases in the low load regime. This is in accordance with the conventional wisdom of materials at low loads, where defect formation is mainly close to the surface, and increasing number of extended defects (dislocations) enhances the strength (hardness) of the material. Formation of extended defects can be explained with geometrically necessary dislocation model [13], which in turn explains the depth dependent hardness variation for the crystalline material.

Typical loading-unloading curves are shown for low (Figs. 2a, b) and high (Figs. 2c, d) loads for a fixed loading-unloading rate of 10 mN.s$^{-1}$ in c-Si substrates. In both cases, 'pop-out' burst in the unloading line (encircled in Figs.. 2a-d) is noticed. The mechanism responsible for the 'pop-out' burst appears to be associated with the formation of second phase(s) [14]. Interestingly, unlike the case for GaN we fail to observe any 'pop-in' burst in the loading line which indicates lack of nucleation and subsequent movement of dislocation sources including lattice atoms [15].

*3.2 Phase identification by Raman studies*

Typical Raman line mappings for c-Si(100) are shown in Fig. 3 for the corner, edge and the central regions of the indentation spot. The Raman peaks corresponding to different reported phases of Si are tabulated (Fig. 3a). Raman mode for Si-I phase of c-Si at 521 cm$^{-1}$ along with zone boundary (ZB) peak at 300 cm$^{-1}$ are shown for the spot at a corner close to the outside region (Fig. 3b). The peak at 521 cm$^{-1}$ is close to reported 520 cm$^{-1}$ optical mode of bulk Si. A small amount of stress may cause this shift. Close to the edge of the indented spot, phases corresponding to c-Si-I (300 and 523 cm$^{-1}$), Si-III (166, 385, 432, and 463 cm$^{-1}$) and Si-XII (182, 351, and 397 cm$^{-1}$) are shown (Fig. 3c). At the centre of the indent (Fig. 3d), crystalline phase corresponding to Si-I almost disappears. Distinct amorphous Si (broad peak around 505 cm$^{-1}$) along with metastable phases of Si-III and Si-XII phases are observed at the center indicating maximum pressure being applied in this zone. The distribution of pressure under the indenter is given by [16],

$$p(r) = \frac{E}{2(1-\nu^2)} \frac{\cosh^{-1}(a/r)}{\tan \Psi}, 0 \leq r \leq a \qquad \ldots\ldots\ldots\ldots\ldots\ldots\ldots\ldots\ldots\ldots (1)$$

where $E$ is Young's modulus [130 GPa for Si(100) and 185 GPa for Si(111)], $\nu$ is Poisson's ratio [0.28 for Si(100) and 0.26 for Si(111)] [17], $a$ is the contact radius, and $r$ is the radial coordinate in the surface. With $\Psi$ of 65.3º in the Berkovich indenter, the pressure at the central region (for a spread of ~1 μm in the micro-Raman resolution at the centre) can be calculated (Eqn.1) as ~ 10 GPa (close to the boundary of ~1 μm spot at the centre) to 74 GPa (close to the centre). Thus Si-III and Si-XII phases are also likely to be observed at the edge of the indent spot (Fig. 3c) [10,11]. The estimated pressure requirement of 24 GPa for the amorphization at room temperature (RT) is also provided at the centre of the indent (Fig. 3d) and the near-by regions, as detailed later. Similar Raman spectra for c-Si(111) at the border (Fig. 4a) and center (Fig. 4b) regions of the indentation spot, are also shown for different high pressure phases of Si as tabulated in Fig. 3a. Like Si(100) case, Si-I phase is prominently present in the border region (Fig. 4a) while amorphous phase (480 and 510 $cm^{-1}$) peaks up at the center (Fig. 4b) of the spot.

*3.3 Raman area mapping*

Raman imaging (Fig. 5a; , a grid of 1 μm square area is indicated for the mapping) of c-Si(100) using the spectral region of 256-315 $cm^{-1}$ and 515-545 $cm^{-1}$ shows green and red regions (bright regions in the grayscale), respectively, lying almost outside the indented region indicating Si-I phase of c-Si (301 and 520 $cm^{-1}$). A different contrast along the edge (blue for 256-315 $cm^{-1}$ and blue and green for 515-545 $cm^{-1}$ imaging) on the triangular indent spot shows the presence of Si-I phase inside the indent spot close to boundaries (as also observed in Fig. 3c). Imaging for the spectral range of 154-170 $cm^{-1}$, 370-390 $cm^{-1}$ and 424-448 $cm^{-1}$ shows red (bright in the grayscale) regions lying inside the indented region corresponding to Raman peaks of Si-III, respectively, at 166, 385 and 432 $cm^{-1}$ (Fig. 5b). Similarly, Raman imaging for the spectral range of 176-184 $cm^{-1}$, 330-365 cm$^{-1}$ and 390-401 cm$^{-1}$ shows red (bright in the grayscale) regions lying inside the indented region corresponding to Raman peaks of Si-XII, respectively, at 183, 351 and 397 cm$^{-1}$ (Fig. 5c). Both Si-III and Si-XII phases seem to lie close to each other in the indented space as they are close variants of pressure [10,11]. Amorphous phase (480 and 505 cm$^{-1}$) is also mapped for the spectral region of 458-515 cm$^{-1}$ and observed to be restricted inside the indented spot (Fig. 5d). We must restate here that instead of amorphous phase being at the center alone, where the pressure is calculated [16] to be the maximum, it is distributed in the larger area of the indentation spot where adequate pressure for amorphization of Si is attained. A contour map (Fig. 5e) of regions with different pressure is calculated using Eqn.(1). Analogous area mapping of the c-Si(111) for Si-III (Fig. 6a), Si-XII (Fig. 6b) and amorphous Si (Fig. 6c) is shown with red (bright in the grayscale) regions lying inside the indented region corresponding to Raman peaks of Si phases. A contour map (Fig. 6d) of regions with different pressure is also calculated for Si(111) using Equation 1. We can observe that pressure of amorphization at RT ~24 GPa exists close to the boundary (Figs. 5e and 6d), which explains the Raman area maps in Fig.5(d) and Fig.6(c) for Si(100) and (111), respectively. Presence of Si-I (0-11GPa) phase only in the boundary of the indented region and co-existence of Si-III and Si-XII phases inside the indented region can thus be explained.

## 4. Conclusion

A complete mapping of the phase transformation in crystalline Si induced by indentation is reported. Raman line as well area mapping are used for locating the major phases. The experimental evidences are supported by calculation of pressure contours over the indented region. A comprehensive understanding of the phase distribution under pressure will be useful for various applications of Si in the micromachining and ion beam processing.

**References**

1. Clarke DR, Kroll MC, Kirchner PD, Cook RF, Hockey BJ, Phys Rev Lett 1998;60:2156-2159.

2. Khayyat MM, Banini G.K, Hasko DG., Chaudhri MM, J Phys D : Appl Phys 2003 ;36,1300-1307.

3. Jang JI, Lance MJ, Wen S, Tsui TY, Pharr GM, Acta Materialia 2005;53:1759-1770.

4. Domnich V, Gogotsi Y, In: Handbook of surfaces and interfaces of materials (Ed: H. S. Halwa) New York: Academic Press 2001**;**2:195.

5. Li ZC, Liu L, Wu X, He LL, Xu YB, Materials Science and Engineering A 2002;337:21-24.

6. Dhara S, Das CR, Hsu HC, Raj B, Bhaduri AK, Chen LC, Chen KH, Albert SK Ray A, Appl Phys Lett 2008;92:143114-1-3.

7. Tanikella BV, Somasekhar AH, Sowers AT, Nemanich RJ, Scattergood RO, Appl Phys Lett 1996;69:2870-2872.

8. Patten J, Fesperman R, Kumar S, McSpadden S, Qu J, Lance M, Nemanich R, Huening J, Appl Phys Lett 2003;83:4740-4742.

9. Wu X, Yu J, Ren T, Liu L, Microelectronics Journal 2007;38:87; Carr DW, Evoy S, Sekaric L, Parpia JM, Craighead HG, Appl Phys Lett 1999;75:920-922.

10. Hu JZ, Merkle LD,. Menoni CS, Spain IL, Phys Rev B 1986;34: 4679-4684.

11. Piltz RO, Maclean JR, Clark SJ, Ackland GJ, Hatton PD, Crain J, Phys Rev B 1995;52: 4072-4085.

12. Oliver WC, Pharr GM, J Mater Res 1992;7:1564-1583.

13. Nix WD, Gao H, J Mech Phys Solid 1998;46:411-425.

14. Domnich V, Gogotsi Y, Rev Adv Mater Sci 2002;3:1-36.

15. Gaillard Y, Tromas C, Woirgard J, Phil Mag Lett 2003;83:553-561.

16. Johnson KL, Contact mechanics, Cambridge: Cambridge University Press; 1985, p. 107.

17. Wortman JJ, Evans RA, J Appl Phys 1965;36:153-156.

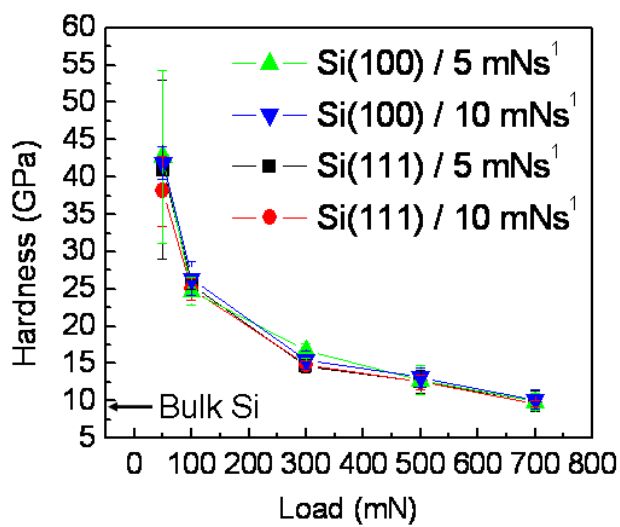

Fig.1. (Color online) Measured hardness of (100) and (111) oriented c-Si substrates using Berkovich indenter for various loads at different loading-unloading rates. Bulk hardness value for Si is indicated by arrow at 8.5 GPa.

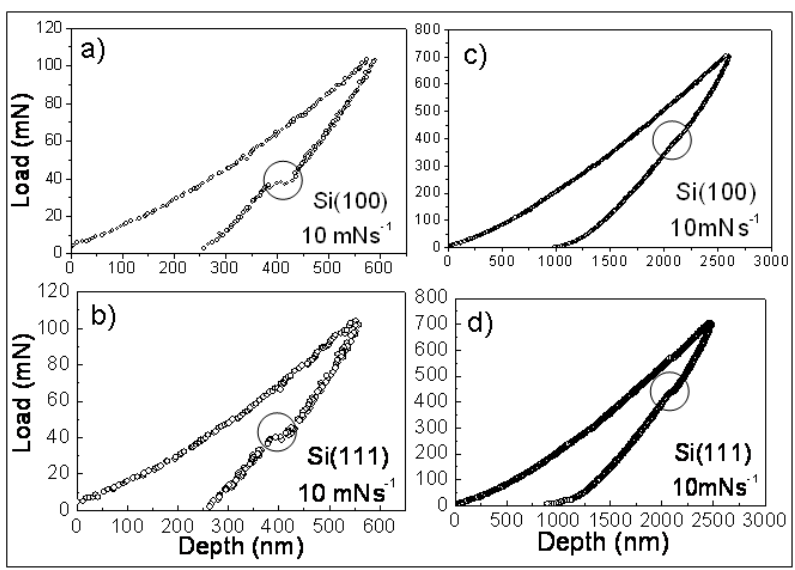

Fig. 2 Typical loading-unloading lines at a fixed loading-unloading rate of 10 mN.s$^{-1}$ for low (100 mN) loads in c-Si substrates with crystalline orientations of a) (100) and b) (111). Corresponding plots are also shown for crystalline orientations of c) (100) and d) (111) with high (700 mN) loads. The encircled regions are called as 'pop-out' burst.

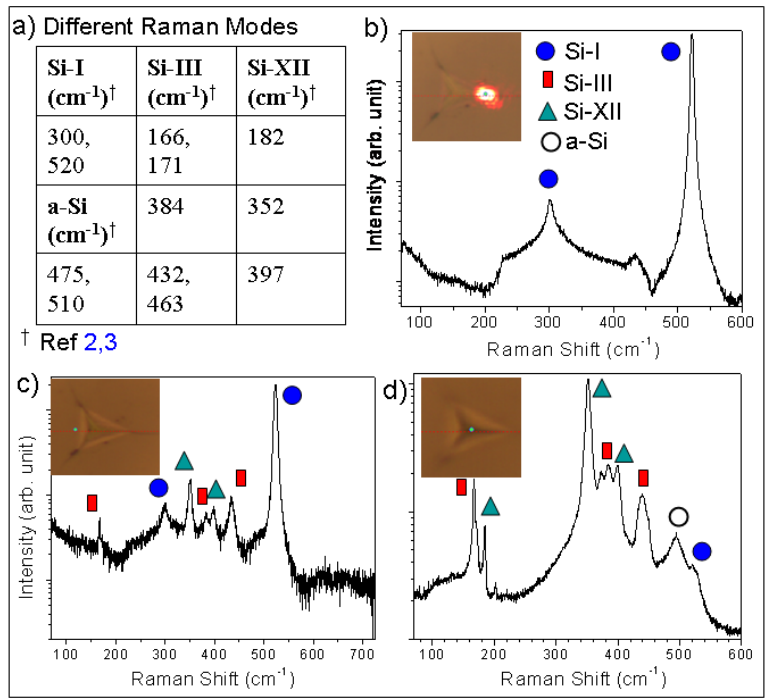

Fig. 3. (Color online) (a) Tabulated values of Raman peak positions for various phases of Si reported in literature. Raman spectra at different regions b) corner, c) edge and (d) centre of indented spot, as shown in the corresponding insets, exhibiting all the phases of Si tabulated in (a).

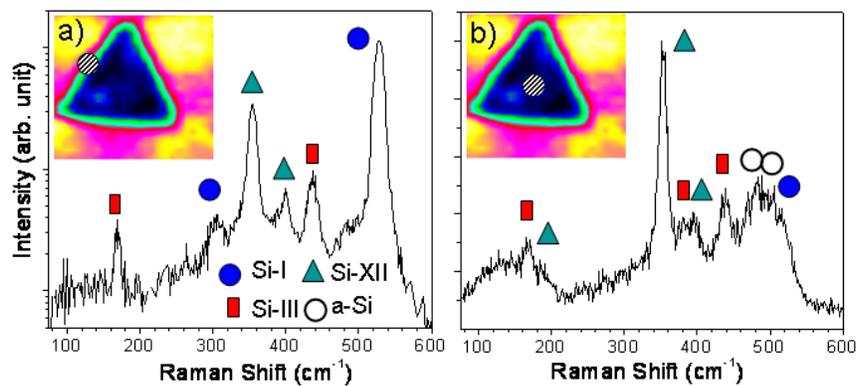

Fig. 4. (Color online) Raman spectra at the a) border and b) center regions, as shown in the corresponding insets of indented spot, exhibiting various phases of Si.

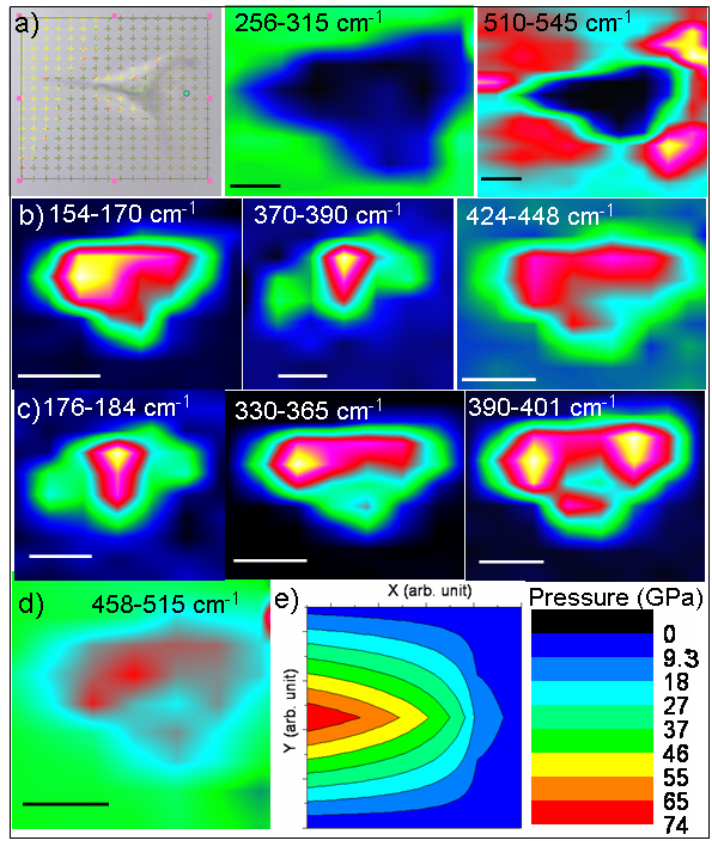

Fig. 5. (Color online) a) Typical grid used for the Raman area maps for c-Si(100). Si-I phases of ZB mode at 300 cm$^{-1}$ and optical mode at 520 cm$^{-1}$ are mapped. (b) Si-III phases around 166, 385, 432 cm$^{-1}$ are mapped (c) Si-XII phases around 182, 352, 397 cm$^{-1}$ are mapped. (d) Amorphous Si is mapped for around 505 cm$^{-1}$ Raman modes. Scale bars are 5 μm. (d) Calculated pressure contour of the indented region.

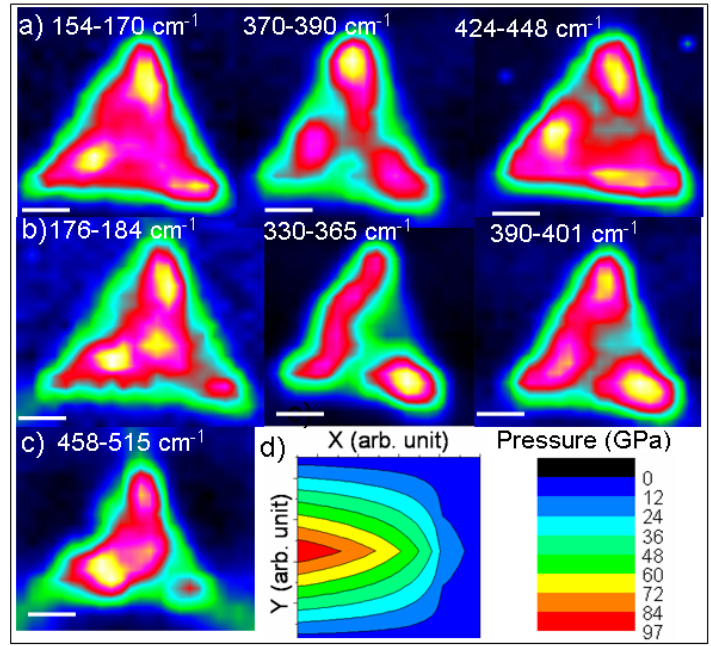

Fig. 6. (Color online) Raman area maps for c-Si(111). (a) Si-III phases around 166, 385, 432 cm$^{-1}$ are mapped (c) Si-XII phases around 182, 352, 397 cm$^{-1}$ are mapped. (d) Amorphous Si is mapped for 480 and 505 cm$^{-1}$ Raman modes. Scale bars are 5 μm. (d) Calculated pressure contour of the indented region.